\documentclass[prm,longbibliography,twocolumn,superscriptaddress,amsmath,amssymb,verbatim]{revtex4-1}

\usepackage{graphicx}
\usepackage{lmodern}
\usepackage{epstopdf}
\usepackage[colorlinks=true,citecolor=blue,linkcolor=blue,urlcolor=blue,bookmarks=true]{hyperref}
\usepackage[sort&compress]{natbib}
\usepackage[normalem]{ulem}
\usepackage{xcolor}

\newcommand{\Cu}{YCu$_3$(OH)$_6$Cl$_3$}
\newcommand{\Cud}{Y$_3$Cu$_9$(OH)$_{18}$[Cl$_8$(OH)]}

\begin{document}

\preprint{APS}

\title{Magnetic ordering of the distorted kagome antiferromagnet Y$_3$Cu$_9$(OH)$_{18}$[Cl$_8$(OH)] prepared via optimal synthesis}

\author{W. Sun}
\affiliation{Fujian Provincial Key Laboratory of Advanced Materials, Department of Materials Science and Engineering, College of Materials, Xiamen University, Xiamen 361005, Fujian Province, People's Republic of China}
\author{T. Arh}
\affiliation{Jo\v{z}ef Stefan Institute, Jamova c.~39, SI-1000 Ljubljana, Slovenia}
\affiliation{Faculty of Mathematics and Physics, University of Ljubljana, Jadranska u.~19, SI-1000 Ljubljana, Slovenia}
\author{M.\,Gomil\v{s}ek}
\affiliation{Jo\v{z}ef Stefan Institute, Jamova c.~39, SI-1000 Ljubljana, Slovenia}
\author{P. Ko\v{z}elj}
\affiliation{Jo\v{z}ef Stefan Institute, Jamova c.~39, SI-1000 Ljubljana, Slovenia}
\affiliation{Faculty of Mathematics and Physics, University of Ljubljana, Jadranska u.~19, SI-1000 Ljubljana, Slovenia}
\author{S. Vrtnik}
\affiliation{Jo\v{z}ef Stefan Institute, Jamova c.~39, SI-1000 Ljubljana, Slovenia}
\author{M. Herak}
\affiliation{Institute of Physics, Bijeni\v{c}ka c.~46, HR-10000 Zagreb, Croatia}
\author{J.-X. Mi}
\affiliation{Fujian Provincial Key Laboratory of Advanced Materials, Department of Materials Science and Engineering, College of Materials, Xiamen University, Xiamen 361005, Fujian Province, People's Republic of China}
\author{A. Zorko}
\email{andrej.zorko@ijs.si}
\affiliation{Jo\v{z}ef Stefan Institute, Jamova c.~39, SI-1000 Ljubljana, Slovenia}
\affiliation{Faculty of Mathematics and Physics, University of Ljubljana, Jadranska u.~19, SI-1000 Ljubljana, Slovenia}

\date{\today}

\begin{abstract} 
Experimental studies of high-purity kagome-lattice antiferromagnets (KAFM) are of great importance in attempting to better understand the predicted enigmatic quantum spin-liquid ground state of the KAFM model. 
However, realizations of this model can rarely evade magnetic ordering at low temperatures due to various perturbations to its dominant isotropic exchange interactions.
Such a situation is for example encountered due to sizable Dzyaloshinskii-Moriya magnetic anisotropy in YCu$_3$(OH)$_6$Cl$_3$, which stands out from other KAFM materials by its perfect crystal structure.  
We find evidence of magnetic ordering also in the distorted sibling compound Y$_3$Cu$_9$(OH)$_{18}$[Cl$_8$(OH)], which has recently been proposed to feature a spin-liquid ground state arising from a spatially anisotropic kagome lattice.
Our findings are based on a combination of bulk susceptibility, specific heat, and magnetic torque measurements that disclose a N\'eel transition temperature of $T_N=11$~K in this material,
which might feature a coexistence of magnetic order and persistent spin dynamics as previously found in YCu$_3$(OH)$_6$Cl$_3$.
Contrary to previous studies of single crystals and powders containing impurity inclusions, we use high-purity single crystals of Y$_3$Cu$_9$(OH)$_{18}$[Cl$_8$(OH)] grown via an optimized hydrothermal synthesis route that minimizes such inclusions. 
This study thus demonstrates that the lack of magnetic ordering in less pure samples of the investigated compound does not originate from the reduced symmetry of spin lattice but is instead of extrinsic origin.       
\end{abstract}

\maketitle

\section{Introduction}
The kagome-lattice antiferromagnet (KAFM) has been one of the most intensively studied spin models in recent years, mostly because of its predicted quantum disordered, yet highly entangled quantum spin-liquid ground state \cite{balents2010spin, savary2017quantum, zhou2017quantum, broholm2020quantum}.
However, in compounds that realize this model, departures from the idealized isotropic nearest-neighbor exchange interactions are inevitable and can crucially impact the magnetic ground state.
For example, the archetypal KAFM material herbertsmithite \cite{norman2016herbertsmithite} is hampered by various perturbations, including substantial intersite ion mixing \cite{de2008magnetic, olariu200817, freedman2010site}, reduction of crystal symmetry \cite{zorko2017symmetry,laurita2019evidence}, and sizable magnetic anisotropy \cite{zorko2008dzyaloshinsky}.
Due to these, likely intertwined, perturbing effects the true origin of its observed gapless spin-liquid ground state \cite{khuntia2020gapless} is still unclear and a deeper understanding of the individual roles of these perturbations remains one of the main objectives in the field.
The problem of structural disorder is particularly notorious, as defects are expected to substantially perturb local spin correlations of the KAFM \cite{dommange2003static, poilblanc2010impurity}, but can also be employed as in-situ probes of the host state \cite{gomilsek2019kondo}.

\begin{table*}[t]
\caption{Optimal recipes for the synthesis of \Cud~single crystals (SC) and powders.
\label{tab1}}
\begin{ruledtabular}
\begin{tabular}{c c|c c c c c c}
Method & Form & LiOH$\,\cdot\,$H$_2$O & Y(NO$_3$)$_3\,\cdot\, 6$H$_2$O & YCl$_3\cdot 6$H$_2$O & CuCl$_2\cdot 2$H$_2$O & Cu$_2$(OH)$_2$CO$_3$ & H$_2$O  \\
\hline
hydrothermal & SC & $1.00$\,g  &  $1.95$\,g &  /  &  $2.65$\,g  & / & $10.00$\,g\\
solid state & powder & / & / & $0.61$\,g & / & $0.66$\,g & / \\
\end{tabular}
\end{ruledtabular}
\end{table*}

The recently synthesized KAFM material \Cu~appears to be closer to the perfect kagome model in a number of aspects \cite{sun2016perfect}. 
Like in herbertsmithite, the nearest-neighbor exchange interactions  $J=82$\,K are by far the dominant isotropic interactions in this compound \cite{arh2020origin}, yet any structural imperfections are absent \cite{sun2016perfect, berthelemy2019local}.
It nevertheless exhibits a long-range ordered 120$^\circ$ ($q=0$) magnetic structure \cite{ zorko2019negative} below $T_N = 12$~K \cite{arh2020origin,  berthelemy2019local, zorko2019negative, zorko2019YCu3muon}, which is stabilized by extremely large out-of-plane Dzyalsohinskii-Moriya (DM) exchange anisotropy, $D/J = 0.25$ \cite{arh2020origin}.
Interestingly, persistent spin dynamics was found to coexist with this order even at temperatures as low as $T/T_N=1/300$ \cite{zorko2019YCu3muon}.
It is predicted that introducing lattice distortions should destabilize the magnetic order on the kagome lattice \cite{coker2021effects} and potentially lead to a spin-liquid ground state.
Such an effect could indeed be at play in the distorted sibling compound \Cud~with two inequivalent copper sites on the kagome lattice \cite{puphal2017strong}. 
In fact, even though this compound was initially proposed to magnetically order around 2\,K where a pronounced peak in specific heat was observed \cite{puphal2017strong}, a later study detected dynamical internal magnetic fields even at 20\,mK and thus suggested a disordered ground state \cite{berthelemy2019local}. 
It was further proposed that the observed specific-heat peak might be an intrinsic feature of the spin liquid in the KAFM model \cite{berthelemy2019local}. 
However, there is a potential caveat to these conclusions.
Namely, the intensity of the peak in specific-heat at 2 K is highly sample dependent and the magnetic ground state is spatially inhomogeneous \cite{berthelemy2019local}. 
Furthermore, synthesizing this compound at a high level of purity has so far proven to be extremely challenging \cite{puphal2017strong, berthelemy2019local}.
The presence of parasitic clinoatacamite impurities is regularly encountered in both \Cud~and \Cu~samples because they both have quite similar octahedral [Cu(OH)$_4$Cl$_2$] layers as clinoatacamite, with a lattice mismatch of only 1.4\%, which gives rise to epitaxial intergrowth and surface coating between these phases \cite{chen2020quantum}. 

Here we exploit an optimized synthesis route for growing high quality  \Cud~single crystals.
These have the same crystal structure as previously reported crystal samples \cite{puphal2017strong}, but possess much fewer magnetic impurities. 
The magnetic behavior of these optically transparent crystals is very differently from the behavior of previously grown translucent crystals, which feature dark inclusions.
By performing bulk susceptibility, specific heat and magnetic torque measurements, we show that the pure, transparent crystals actually undergo a magnetic ordering transition at $T_N = 11$\,K, much like the non-distorted sibling compound \Cu.
The effect of structural distortion in \Cud~away from a perfectly symmetric kagome lattice apparently does not favor a spin-liquid ground state. 
However, additional disorder due to impurity inclusions in less pure crystals does seem to suppress long-range ordering.
Our study thus elucidates the problem of impurities in KAFM materials from a new perspective and shows that neglecting these can lead to erroneous conclusions about the fragile magnetic ground state of this paradigmatic geometrically frustrated model.  

\section{Synthesis}
\begin{figure}[b]
\includegraphics[trim = 0mm 0mm 0mm 0mm, clip, width=1\linewidth]{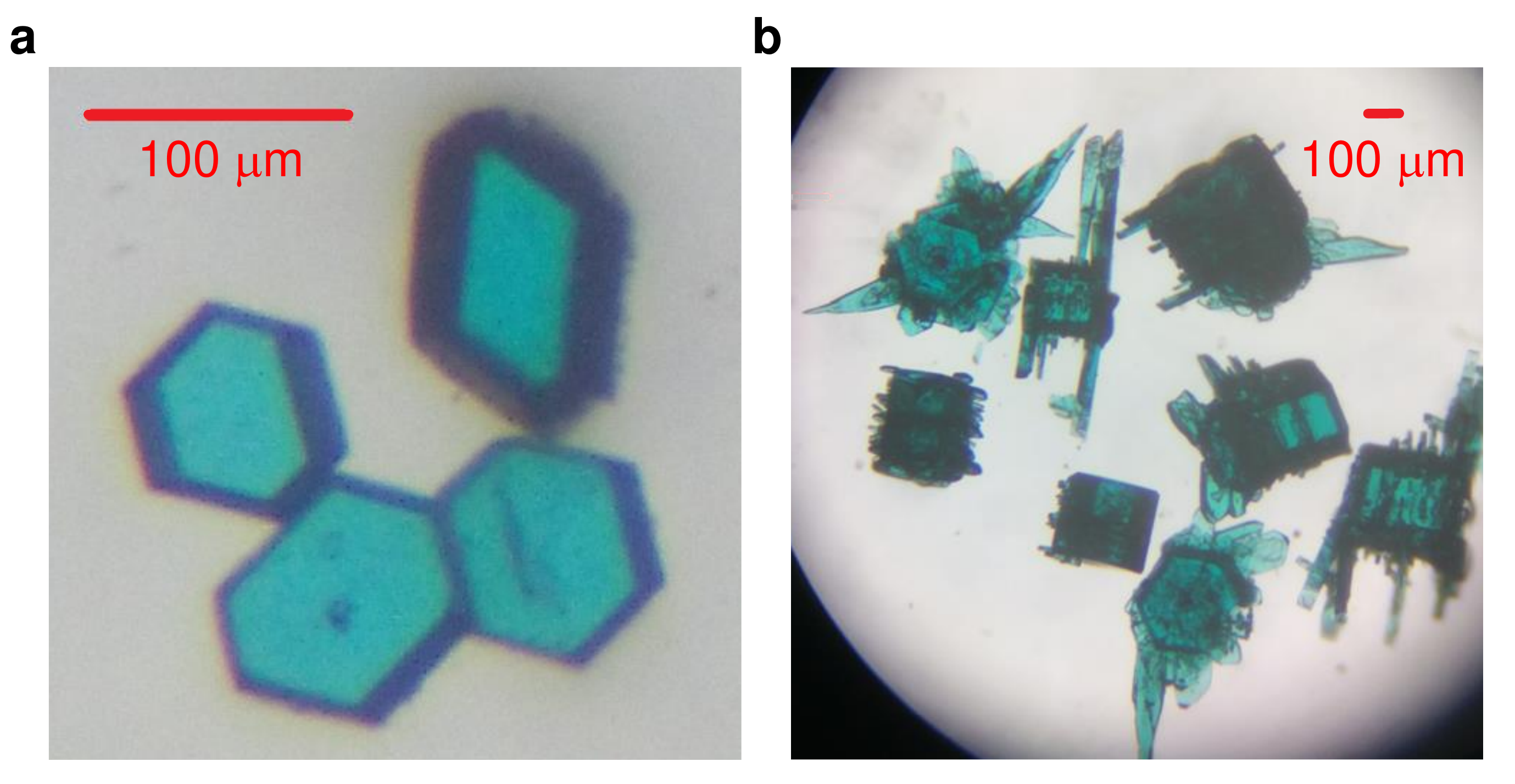}
\caption{(a) Optical image of transparent single crystals of \Cud~with black inclusions of CuO in the lower two crystals and no visible inclusions in the upper two.
The crystals are hexagonally shaped with the crystallographic $c$ axis perpendicular to the surface of the platelets.  
(b) Larger translucent aggregates of crystals possess abundant inclusions of CuO and clinoatacamite Cu$_2$(OH)$_3$Cl.}
\label{fig1}
\end{figure}

The initially reported synthesis of single crystals of \Cud~was based on the hydrothermal method \cite{puphal2017strong}, while powders of this compound were later synthesized via a solid-state reaction \cite{berthelemy2019local}.
We have explored both of these synthesis routes in detail.
With the hydrothermal route the best results were obtained when using $1.00$\,g of LiOH$\cdot$H$_2$O ($23.83$\,mmol), $1.95$\,g of Y(NO$_3$)$_3\cdot6
$H$_2$O ($5.01$\,mmol), $2.65$\,g of CuCl$_2\cdot2$H$_2$O ($15.54$\,mmol) and $10.00$\,g of distilled water ($10.00$\,mmol) as starting materials (see Table\,\ref{tab1}).
These were sequentially added into a $25$-mL Teflon-lined stainless steel autoclave without stirring, i.e., without homogenization. 
The autoclaves were then heated in a furnace and kept at $463$\,K for 3 days. 
The solid products consisted of \Cud~single crystals of maximum dimension $\sim$150\,$\mu$m, where some of them were transparent and contained no visible inclusions, while the majority of them contained dark inclusions (Fig.\,\ref{fig1}a).
Aggregates of large translucent blue crystals (up to $\sim$1 mm in size) that were also grown were found to always contain dark inclusions (Fig.\,\ref{fig1}b).
These are due to clinoatacamite Cu$_2$(OH)$_3$Cl and CuO impurities, as also found in previously grown crystals \cite{puphal2017strong}.
Such impurities appear not only on the surface but also inside the crystals, e.g., on  cleavage planes (Fig.\ref{fig1}b). 
We note that washing the products with distilled water converts them to blue translucent crystals, which also include clinoatacamite and CuO impurities.
Unfortunately, only CuO particles adsorbed on the crystal surface can be cleaned in an ultrasonic bath with pure alcohol.
Therefore, hand picking under an optical microscope was required to separate pure transparent single crystals from translucent ones. 

\begin{table*}[t]
\caption{Crystal data and structural refinements of \Cud~in the subcell and supercell. Details are available in Ref.\,\cite{sup}.
\label{tab2}}
\begin{ruledtabular}
\begin{tabular}{l l l}
& Subcell &	Supercell   \\
\hline
Formula sum, weight, $Z$ &	Cl$_{2.65}$Cu$_3$H$_{6.35}$O$_{6.35}$Y, 481.47, 1 & Cl$_8$Cu$_9$H$_{19}$O$_{19}$Y$_3$, 1445.34, 3 \\
Size (mm), color & \multicolumn{2}{c}{$0.15 \times 0.10 \times 0.10$, blue}\\
System, space group & Trigonal, $P\bar{3}$ (No.\,147) & Trigonal, $R\bar{3}$  (No.\,148)\\
Cell $a, c$ (\AA) & 6.6644(9), 5.7293 (11) & 11.543(3), 17.188(4)\\
Volume (\AA$^3$) & 220.37(7) & 1983.3(11)\\
Target, $\lambda$ (\AA), $T$ (K) & \multicolumn{2}{c}{Mo K$\alpha$, 0.71073, 173}\\
$\mu$ (mm$^{-1}$), $F$(000), $Dx$ (kg m$^{-3}$) & 14.452, 228, 3628 & 14.46, 2055, 3630\\
$hkl$ ranges & $-6$ to 8, $-8$ to 4, $-7$ to 7 & $-6$ to 15, $-15$ to 12, $-22$ to 22\\
$2\theta_{\rm max}$($^\circ$), $N_{\rm par}$ & 56.40, 30 & 56.40, 70\\
$R_{int}$, $R_1$, w$R_2$, $S$ & 0.030, 0.023, 0.062, 1.012 & 0.037, 0.035, 0.094, 1.07\\
$N_{\rm ref}$, for $I > 2\sigma(I)$ & 351, 332 & 1044, 844\\
$\Delta \rho_{\rm max}$, $\Delta \rho_{\rm min}$ (e \AA$^{-3}$) & $-0.404$, 0.486 & 1.13, $-0.84$\\
\end{tabular}
\end{ruledtabular}
\end{table*}
\begin{table*}[t]
\caption{Fractional atomic coordinates and equivalent isotropic displacement parameters (in \AA$^2$) in the supercell of \Cud.
Refinement with an unconstrained Y occupancy yields an occupancy sum of 2.998(42), in excellent agreement with the theoretical value of 3.
All other sites are fully occupied within the experimental uncertainty, thus demonstrating perfect stoichiometry of our transparent single crystals.
The hydrogen sites from Ref.\,\cite{puphal2017strong} were not refined and are omitted from this table. 
\label{tab3}}
\begin{ruledtabular}
\begin{tabular}{c c c c c c c c}
Atom &	Wyck. &	Site & $x$ &	$y$ &	$z$ &	$U_{\rm iso}$  &	Occ.($<1$)\\
\hline
Y1 & $6c$ & 3. & 0.0000 & 0.0000 & 0.12805(7) & 0.0116(3) & 0.953(5) \\
Y2 & $6c$ & 3. & 0.0000 & 0.0000 & 0.2042(13) & 0.0116(3) & 0.047(5)\\
Y3 & $3b$ & $\bar{3}$. & 0.3333 & 0.6667 & 0.1667 & 0.0120(11) & 0.92(2)\\
Y4 & $6c$ & 3. & 0.3333 & 0.6667 & 0.126(4) & 0.0120(11) & 0.039(10) \\
Cu1 & $18f$ & 1 & 0.15896(11) & 0.32977(8) & 0.16321(5) & 0.0120(2) & \\
Cu2 & $9d$ & $\bar{1}$ & 0.3333 & 0.1667 & 0.1667 & 0.0124(3) & \\
Cl1 & $18f$ & 1 & 0.33554(16) & 0.33144(16) & 0.04917(7) & 0.0218(3) & \\
Cl2 & $6c$ & 3. & 0.3333 & 0.6667 & 0.32845(15) & 0.0210(6) & \\
O1 & $18f$ & 1 & 0.1723(4) & 0.1749(5) & 0.2018(2) & 0.0119(9) & \\
O2 & $18f$ & 1 & $-0.0054(7)$ & 0.1965(5) & 0.1093(2) & 0.0132(9) & \\
O3 & $18f$ & 1 & 0.3254(8) & 0.4701(5) & 0.2098(2) & 0.0134(9) & \\
O4 & $3a$ & $\bar{3}$. & 0.0000 & 0.0000 & 0.0000 & 0.037(3) & \\
\end{tabular}
\end{ruledtabular}
\end{table*}

We also performed a low-temperature solid-state reaction to synthesize high-purity powder samples, i.e., nano-particles of \Cud.
Mixing $0.66$\,g of Cu$_2$(OH)$_2$CO$_3$ ($3.00$\,mmol) and $0.61$\,g of YCl$_3\cdot 6$H$_2$O ($2.00$\,mmol) was found to yield the best results (see Table\,\ref{tab1}) when the starting materials were thoroughly homogenized by vigorous stirring and transferred into an autoclave for heating in a furnace at 463\,K for 3 days.
In contrast, when using a heterogeneous starting mixture and a small amount of LiOH$\cdot$H$_2$O, the solid state reaction method led to the growth of \Cu~\cite{sun2016perfect} single crystals, demonstrating that the two sister compounds can be synthesized from the same starting materials.
A crucial factor for determining the formation of these two compounds is water -- when distilled water was added into the starting materials, \Cud~with clinoatacamite and CuO impurities was obtained. 
Thus, we have found that the hydrothermal method clearly prefers \Cud~over \Cu~and is the optimal route for producing high-quality single crystals of \Cud.
For these, the inclusions of clinoatacamite and CuO impurities can be avoided by hand-picking transparent crystals, 
while such impurities can never be completely avoided in nano-crystallites.

\section{Crystal structure}
\begin{figure}[b]
\includegraphics[trim = 0mm 0mm 0mm 0mm, clip, width=1\linewidth]{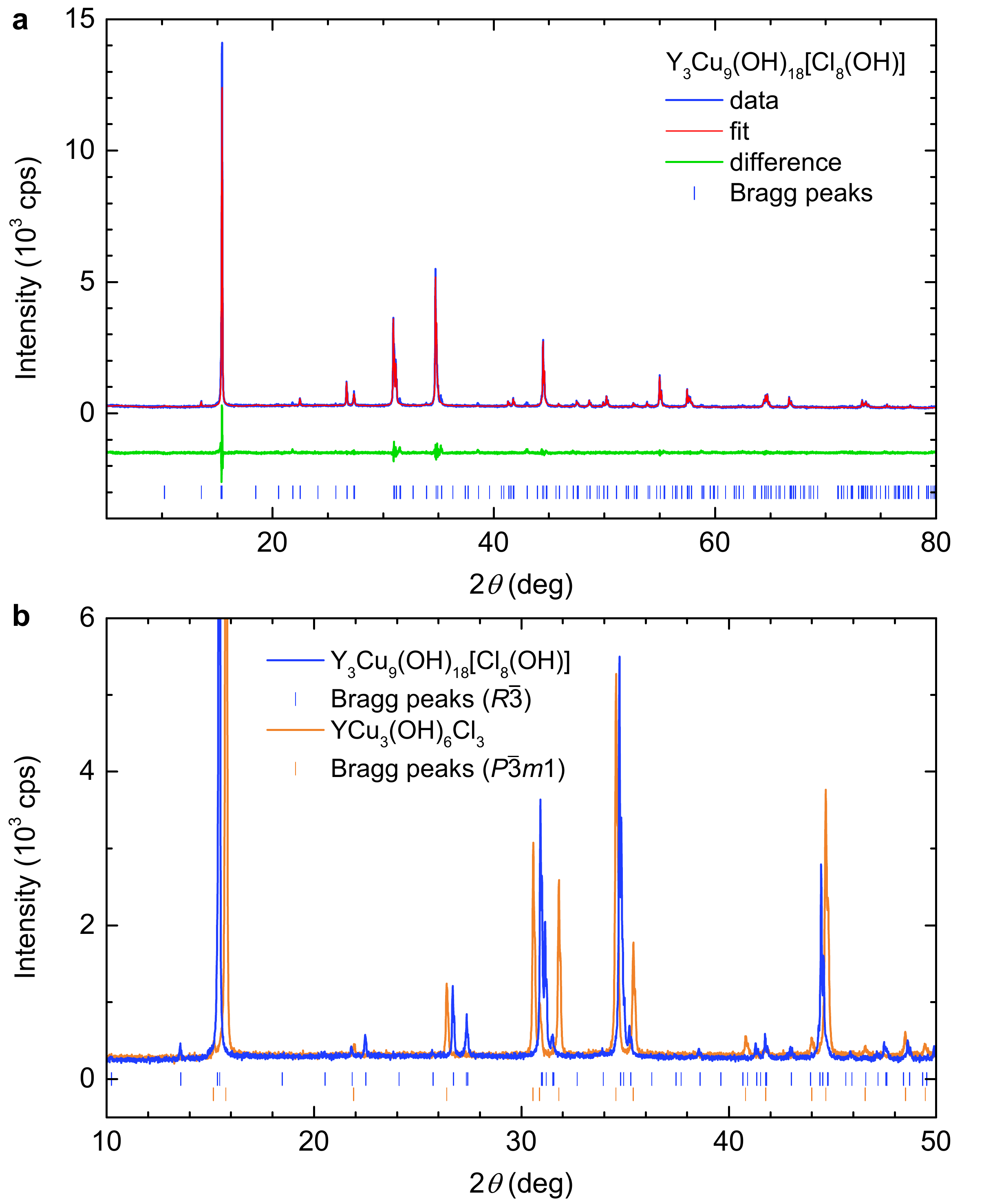}
\caption{(a) Rietveld refinement of the PXRD pattern of \Cud~pulverized high-purity single crystals obtained via the hydrothermal method. 
(b) Comparison of the PXRD patterns of \Cud~(space group $R\bar{3}$, No.\,148) and \Cu~(space group $P\bar{3}m1$, No.\,164), the theoretical positions of Bragg reflections for both compounds are indicated by vertical bars.}
\label{fig2}
\end{figure}

\begin{figure*}[t]
\includegraphics[trim = 0mm 0mm 0mm 0mm, clip, width=1\linewidth]{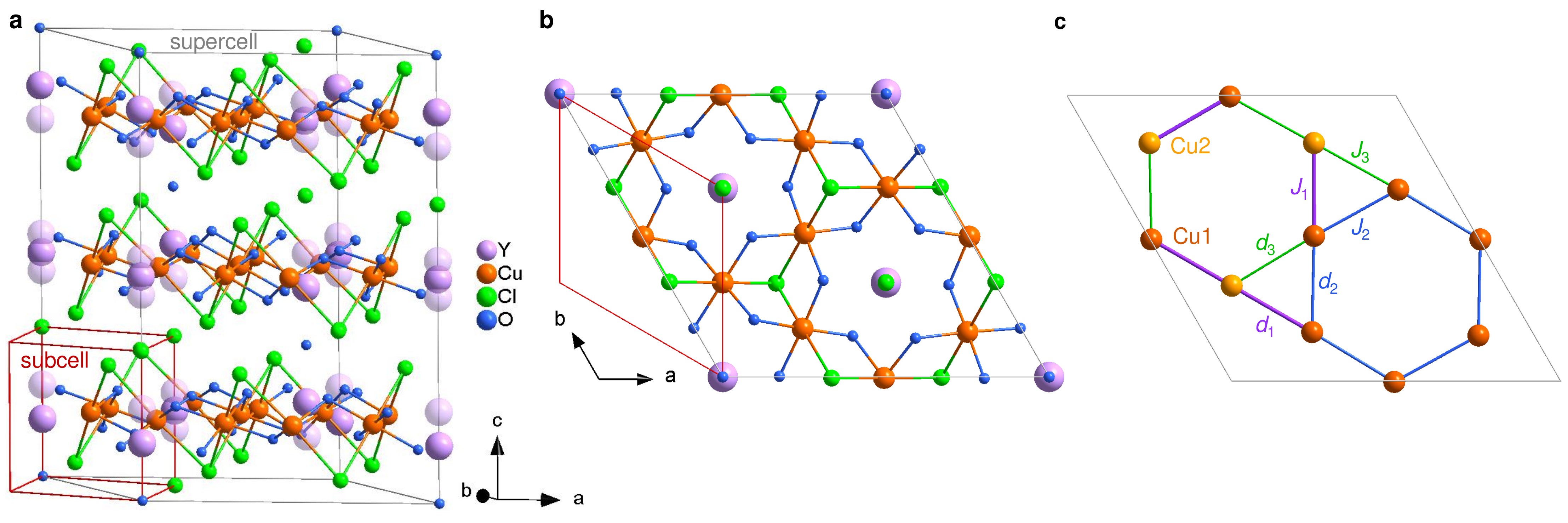}
\caption{(a) The layered crystal structure of \Cud. 
The transparent violet spheres correspond to Y2 and Y4 sites with very low occupancy, which is due to Y$^{3+}$ disorder. 
The edges of the subcell are marked in red and the edges of the supercell in gray. 
(b) In the crystallographic $ab$ planes the magnetic Cu$^{2+}$ ions forming [CuO$_4$Cl$_2$] octahedra are arranged into a kagome lattice, while non-magnetic Y$^{3+}$ ions fill the hexagonal voids. 
(c) The kagome lattice is distorted because of two inequivalent copper sites Cu1 and Cu2, which leads to three different exchange interactions ($J_1$, $J_2$, and $J_3$) between neighboring Cu$^{2+}$ sites.}
\label{fig3}
\end{figure*}

\begin{table*}[t]
\caption{Fractional atomic coordinates and equivalent isotropic displacement parameters (in \AA$^2$) in the subcell of \Cud.
Refinement with an unconstrained Y occupancy yields an occupancy sum of 1.003(10), in excellent agreement with the theoretical value of 1.
\label{tab4}}
\begin{ruledtabular}
\begin{tabular}{c c c c c c c c}
Atom &	Wyck. &	Site & $x$ &	$y$ &	$z$ &	$U_{\rm iso}$ &	Occ.($<1$)\\
\hline
Cu1 & $3e$ & $\bar{1}$ & 0.0000 & 0.5000 & 0.0000 & 0.0150(2) &	\\
Y1 & $2c$ & 3.. & 0.0000 & 0.0000 & 0.1149(4) & 0.0133(4) & 0.352(2)\\
Y2& $1a$ & $\bar{3}$.. & 0.0000 & 0.0000 & 0.0000 & 0.0133(5) & 0.299(4)\\
Cl1 & $2d$ & 3.. & 0.3333 & 0.6667 & 0.3523(2) & 0.0227(3) & \\
Cl2 & $1b$ & $\bar{3}$.. & 0.0000 & 0.0000 & 0.5000 & 0.0257(9) & 0.655(16)\\
O1 & $6g$ & 1 & 0.8117(4) & 0.1887(4) & 0.1339(5) & 0.0298(6) & \\
O2 & $1b$ & $\bar{3}$.. & 0.0000 & 0.0000 & 0.5000 & 0.0257(9) & 0.345(16)\\
\end{tabular}
\end{ruledtabular}
\end{table*}
Powder X-ray diffraction (PXRD) patterns were recorded at 293 K on a Bruker D8 Advance diffractometer with Cu K$\alpha$ radiation ($\lambda = 1.54056$\,\AA), operated at 40\,kV and 40\,mA. 
Single-crystal XRD data were collected at 173 K on a Bruker Apex CCD diffractometer equipped with graphite-monochromatic Mo K$\alpha$ radiation ($\lambda = 0.71073$\,\AA), operated at 45\,kV and 30\,mA. 
Data processing was performed with the SAINT software package. 
The crystal structure of \Cud~was solved by a direct method and refined by full matrix least-squares methods by using the SHELXL crystallographic software package \cite{sheldrick2015crystal}.
The solution was obtained from the PXRD dataset (Fig.\,\ref{fig2}a) of pulverized, hand-selected, high-purity single crystals from hydrothermal synthesis and is broadly consistent with previous reports \cite{puphal2017strong, berthelemy2019local}.
It corresponds to a rhombohedral lattice (Fig.\,\ref{fig3}) with the space group $R\bar{3}$ (referred to as the supercell hereafter; see Table\,\ref{tab2} for details of the refinement and Table\,\ref{tab3} for atomic coordinates and equivalent isotropic displacements). 
The lattice parameters are $a = 11.545(7)$\,\AA~and $c = 17.188(4)$\,\AA.


In order to establish the relation with the much smaller unit cell of the sibling compound \Cu, a suitable subcell of \Cud~was also determined. 
This subcell (red edges in Fig.\,\ref{fig3}a) with parameters $a = 6.7308(6)$\,\AA~and $c = 5.6406(5)$\,\AA~(Tables\,\ref{tab2} and \ref{tab4}) corresponds to the space group $P\bar{3}$.
These cell parameters are very similar to those of \Cu~\cite{sun2016perfect} and other lanthanide cuprates LnCu$_3$(OH)$_6$Cl$_3$ (Ln = Sm, Gd, Nd) \cite{sun2017strong}, however, the subcell symmetry is lower then in these compounds, where it is described by the space group $P\bar{3}m1$.
We note that clear differences in both reflection intensities and Bragg positions between \Cud~and other above-mentioned lanthanide cuprates (Fig.\,\ref{fig2}b) reveal a distinct crystal structure of the former compound, which is the only one to possess a supercell.
As this supercell can be derived from the unit cell of \Cu~-- the $c$ axis is enlarged 3-times and the $a$ axis $\sqrt{3}$-times (see Table\,\ref{tab2} and Fig.\,\ref{fig3}a) -- the two crystal structures bear important similarities.  
They both consist of six-fold coordinated, elongated [CuO$_4$Cl$_2$] octahedra that share O--Cl edges and form two dimensional kagome layers of spin-1/2 Cu$^{2+}$ ions.
The hexagonal holes in the kagome lattice are occupied by diamagnetic Y$^{3+}$ ions. 
We note that our structural refinements did not detect any anti-site mixing between Cu$^{2+}$ and Y$^{3+}$ ions in \Cud, just like previously reported for \Cu~\cite{sun2016perfect}.
This is in sharp contrast to the majority of other Cu-based kagome compounds, like the paradigmatic herbertsmithite \cite{norman2016herbertsmithite}, Zn-brochantite \cite{li2014gapless}, and Zn-doped barlowite \cite{feng2017gapped}, where Zn--Cu intersite disorder amounts to 5--10\% even in the purest samples.


Further comparison of the crystal structures of \Cud~and \Cu~also reveals important structural differences.
The first is a partial replacement of one ninth of all Cl$^-$ ions by hydroxyl groups (OH)$^-$ in the former compound, which are long-range ordered and thus lead to a larger unit cell.
This replacement is suggested already by occupancy refinements in the subcell, yielding the O2/Cl2 ratio of 0.345(16)/0.655(16), which is in excellent agreement with the expected ratio 0.333/0.667.
Furthermore, contrary to \Cu~where all Y$^{3+}$ ions reside at inversion centers and adopt a ``6+2" coordination in [YO$_6$Cl$_2$] polyhedra, only one third of Y$^{3+}$ ions form such polyhedra in \Cud, whereas the remaining Y$^{3+}$ ions are located at non-centered sites where they form [YO$_6$Cl(OH)] polyhedra.
Contrary to previous modeling \cite{puphal2017strong, berthelemy2019local}, our refinements suggest Y-site disorder of the order of a few percents on both majorly occupied sites Y1 and Y3 (Table\,\ref{tab3}), as removing the Y-disorder from the model results in somewhat worse refinements; $R_1$ increases from 0.035 to 0.037, w$R_2$ from 0.094 to 0.098, and the goodness of fit $S$ from 1.07 to 1.11. 
The last major difference between \Cud~and \Cu~is that there is only one crystallographically distinct Cu site in \Cu~\cite{sun2016perfect}, while two inequivalent Cu sites are present in \Cud.
As no symmetry restriction on the $c$ fractional coordinates exists for these two sites, the kagome planes get slightly buckled (Cu1--Cu2 displacement of $\pm 0.059$\,\AA), while they are perfectly flat in \Cu.

We note that a recent neutron diffraction study suggested that the partial substitution of the Cl$^{-}$ ions in \Cud~is due to O$^{2-}$ ions rather than (OH)$^-$ groups, which would imply the composition Y$_3$Cu$_9$(OH)$_{18}$[OCl$_8$] \cite{berthelemy2019local}.
Our XRD data does not allow us to refine the positions of the hydrogen atoms, however, such a chemical formula would lead to excess charge.

\section{Magnetism}
\subsection{Bulk Susceptibility}
\begin{figure*}[t]
\includegraphics[trim = 0mm 0mm 0mm 0mm, clip, width=1\linewidth]{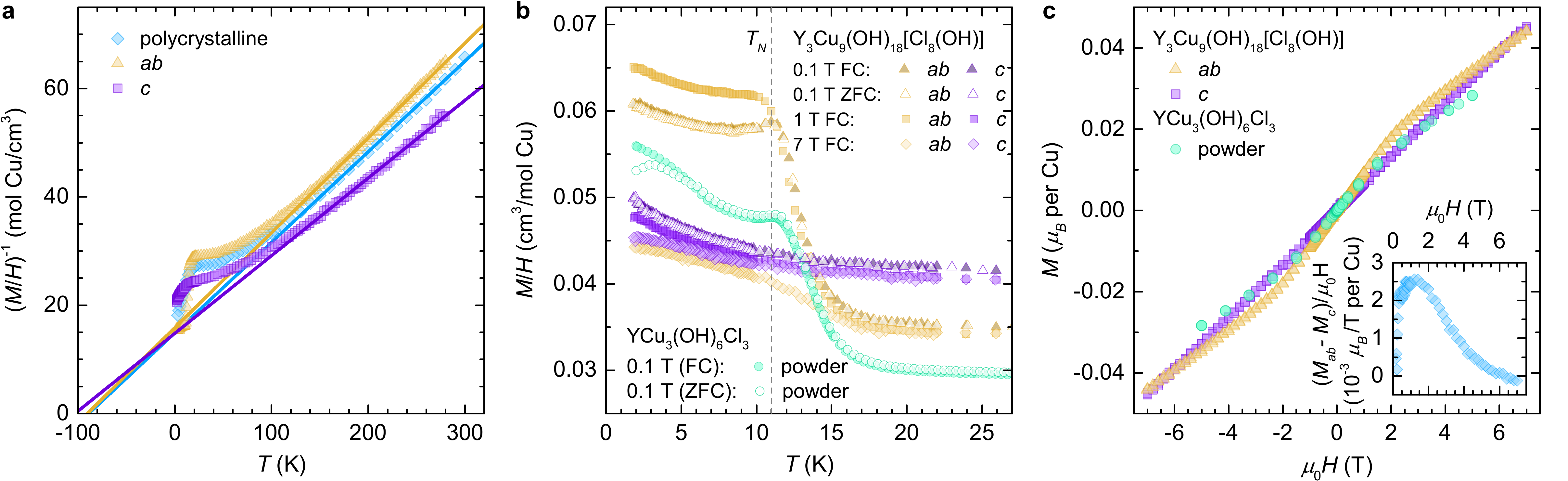}
\caption{(a) Inverse magnetic susceptibility ($M$ stands for magnetization) in an applied field of $\mu_0H= 1$\,T for a polycrystalline sample of \Cud~containing 17\,mg of hand-selected single crystals and for a 0.32-mg single crystal.
$c$ corresponds to the direction of the magnetic field parallel with the high-symmetry crystallographic $c$ axis (i.e., perpendicular to the kagome $ab$ planes), while $ab$ denotes a random direction within the kagome plane.
The solid lines are fits with the Curie-Weiss model for $T>150$\,K.
(b) The low-temperature magnetic susceptibility measured in several magnetic fields and compared to measurements on a \Cu~powder sample from Ref.\,\cite{zorko2019YCu3muon}.
The vertical line highlights the N\'eel temperature $T_N = 11$\,K of \Cud, which is only slightly smaller than the N\'eel temperature $T_N = 12$\,K of \Cu.
(c) The field dependence of magnetization at 2\,K.
Inset shows the difference between the in-plane and the out-of-plane susceptibility at 2\,K as a function of the applied field.}
\label{fig4}
\end{figure*}   

In order to investigate the intrinsic magnetism of \Cud~we first look at the bulk magnetic response of hand-selected transparent pure single crystals. 
The measurements were done on a Quantum Design MPMS3 SQUID magnetometer at temperatures between 1.8 and 300\,K and magnetic fields up to 7\,T.
The magnetic susceptibility $\chi=(M/H)$, where $M$ denotes the measured magnetization in an applied magnetic field $H$, of a 17-mg polycrystalline sample of several tens of single crystals nicely corresponds to a powder average of measurements made on a selected 0.32-mg single crystal for the magnetic field applied along the high-symmetry crystallographic $c$ axis, i.e., perpendicular to the kagome planes, and perpendicular to this axis, i.e., within to the kagome planes, as shown in Fig.\,\ref{fig4}a.
Several other transparent single crystals were also inspected and were found to behave very similarly to the selected crystal. 

From the Curie-Weiss fit of the magnetic susceptibility, $\chi=C/(T-\theta)$, at temperatures above 150\,K, we obtain the Weiss temperatures $\theta_{ab} = - 91(3)$\,K, $\theta_{c} = - 103(3)$\,K, and $\theta_{\rm poly} = - 89(3)$\,K for a magnetic field applied within the kagome lattice, perpendicular to the kagome lattice, and for the polycrystalline sample, respectively.
The corresponding Curie constants $C_{ab} = 5.68(11)$\,cm$^3$\,K/mol\,Cu, $C_{c} = 6.93(13)$\,cm$^3$\,K/mol\,Cu, and $C_{\rm poly} = 5.99(12)$\,cm$^3$\,K/mol\,Cu yield the $g$ factors $g_{ab} = 2.20(2)$, $g_{c} = 2.43(3)$, and $g_{\rm poly} = 2.26(2)$.
The single-crystal values are very similar to those initially reported for a translucent single crystal \cite{puphal2017strong}, indicating that high-temperature magnetism of less pure crystals is not significantly affected by impurities.
The Weiss temperature is also very similar to the value $\theta = - 99$\,K found in the sibling compound \Cu~\cite{arh2020origin}, suggesting that the average nearest-neighbor exchange interaction in the distorted kagome lattice of \Cud~is not far from the single nearest-neighbor interaction in the more symmetric \Cu. 

Below approximately 150\,K the magnetic susceptibility of \Cud~starts to depart from the Curie-Weiss model due to the onset of spin correlations, however, the ratio $\chi_{ab}/\chi_c=g_{ab}/g_c<1$ remains almost unchanged down to about 17\,K.    
Below this temperature, $\chi_{ab}$ starts increasing much more profoundly than $\chi_c$ and exceeds it below 14\,K, except in large applied magnetic fields (Fig.\,\ref{fig4}b).
These drastically suppress the increase of $\chi_{ab}$ at the lowest temperatures, while they have only a marginal effect of $\chi_c$ (Fig.\,\ref{fig4}b).
At 11\,K, a pronounced peak is observed in $\chi_{ab}$, which is very similar to the one found in the magnetic susceptibility of \Cu~powder at its N\'eel ordering temperature of $T_N=12$\,K (Fig.\,\ref{fig4}b) \cite{zorko2019YCu3muon}.
The corresponding sharp increase of the susceptibility of \Cu~in 0.1\,T just above $T_N$ (0.017\,cm$^3$/mol\,Cu) is almost the same as in \Cud~(0.015\,cm$^3$/mol\,Cu if powder averaging the data).  
This strongly suggests that long-range magnetic order develops in bulk of our \Cud~crystals below the N\'eel temperature of $T_N = 11$\,K, as the entire \Cu~sample was shown to undergo magnetic ordering \cite{zorko2019YCu3muon}. 
This is it stark contrast to previous investigations, which failed to detect static magnetism in polycrystalline \Cud~\cite{berthelemy2019local}. 
Furthermore, its low-temperature magnetic response strongly resembles that of the related compound EuCu$_3$(OH)$_{6}$Cl$_3$, where a similar change of anisotropy in magnetic susceptibility was observed at very similar temperatures and was also ascribed to long-range magnetic ordering \cite{puphal2018kagome}.
Contrary to the latter case, we do not observe any hysteresis in the magnetization curve at 2\,K (Fig.\,\ref{fig4}c), so there is no ferromagnetic ordered component neither within nor perpendicular to the kagome planes.
We also note that no difference is found between zero-field-cooled (ZFC) and field-cooled (FC) measurements in \Cud~down to 1.8\,K even in small fields, contrary to \Cu, where a small ZFC/FC splitting was observed below 6.5\,K (Fig.\,\ref{fig4}b) and was attributed to a tiny fraction (0.1\%) of parasitic clinoatacamite \cite{zorko2019YCu3muon}.
Our transparent \Cud~single crystals thus evidently lack even trace amounts of clinoatacamite impurities, even though these were found to be present in all previously synthesized translucent crystals and powders \cite{puphal2017strong,berthelemy2019local}. 

\subsection{Specific Heat}
\begin{figure*}[t]
\includegraphics[trim = 0mm 0mm 0mm 0mm, clip, width=1\linewidth]{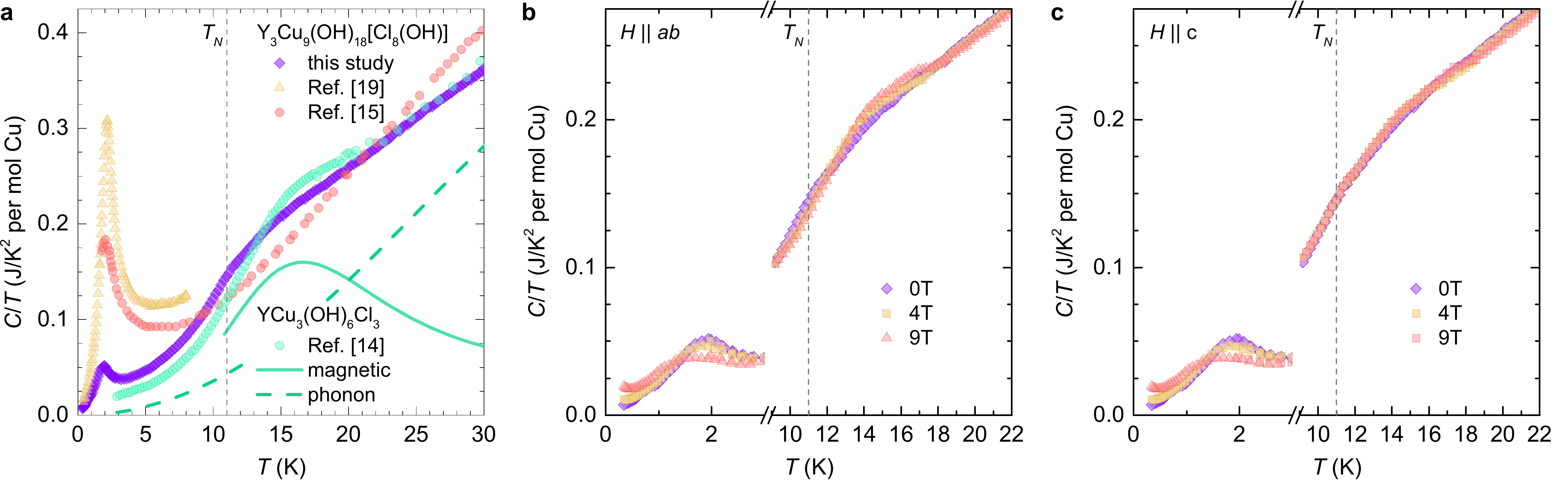}
\caption{(a) 
The temperature dependence of the specific heat of a 0.32-mg \Cud~single crystal in zero magnetic field compared to previously published data by Puphal {\it et al.} \cite{puphal2017strong} and Berth\'elemy {\it et al.} \cite{berthelemy2019local}, and to the \Cu~powder data from Ref.\,\cite{arh2020origin}. 
The solid and the dashed lines indicate the magnetic and the phonon contribution to the specific heat in the latter compound, respectively \cite{arh2020origin}.
The field dependence of the specific heat for a field applied (b) within the kagome planes and (c) perpendicular to the kagome planes.
The vertical dashed lines highlight the N\'eel temperature $T_N = 11$\,K.
The inset in (c) shows a clear peak in the temperature derivative of $C/T$ an $T_N$.
}
\label{fig5}
\end{figure*}   

To further investigate the apparent magnetic ordering in \Cud, specific heat measurements were performed on a Quantum Design MPMS instrument at temperatures between 0.34 and 30 K and magnetic fields up to 9~T.
All investigated transparent single crystals show very similar specific-heat behavior.
Typical results on the 0.32-mg sample that was also used for bulk magnetic characterization in the previous subsection are presented in Fig.\,\ref{fig5} where they are compared to the specific heat of \Cu~\cite{arh2020origin}.
In the latter compound, a broad maximum was found in specific heat around 16\,K, i.e., above its N\'eel ordering temperature $T_N=12$\,K (Fig.\,\ref{fig5}a), which was attributed to a build-up of in-plane chiral spin correlations due to sizable DM interaction \cite{arh2020origin}.
This feature is much less pronounced in \Cud~(Fig.\,\ref{fig5}a), where it appears to be broadened and shifted to slightly lower temperatures.
We assign this to the distortion of the kagome lattice in this compound, which causes a distribution of nearest-neighbor isotropic exchange interactions and DM anisotropies, thus suppressing the build-up of highly symmetric chiral correlations on individual kagome spin triangles above $T_N$.
As a further result of this suppression, the usual anomaly in specific heat at $T_N$, which is absent in \Cu~because the effective number of degrees of freedom involved in long-range ordering is strongly reduced due to the establishment of short-range chiral spin correlations already above the N\'eel temperature \cite{arh2020origin}, becomes more pronounced in \Cud~and thus observable as a tiny bump at $T_N = 11$\,K (Fig.\,\ref{fig5}a). 
It is still extremely small though, as is expected for a strongly two-dimensional spin system \cite{sengupta2003specific}.
The transition at $T_N$ is, however, clearly observed as a peak in the temperature derivative of $C/T$ (insert in Fig.\,\ref{fig5}c).
A magnetic field applied within the kagome planes (Fig.\,\ref{fig5}b) shifts the weight of the specific heat to slightly higher temperatures, suggesting that it stimulates the build-up of spin correlations, while a magnetic field applied perpendicular to the kagome planes shows no such effect (Fig.\,\ref{fig5}c).
Such behavior is in line with the strong increase of the in-plane magnetic susceptibility below 17\,K, while the out-of-plane susceptibility increases much more moderately (Fig.\,\ref{fig4}b).  

Next, we comment on the origin of the low-temperature 2-K peak in specific heat that was initially proposed to be a sign of magnetic ordering \cite{puphal2017strong} and later suggested to rather be an intrinsic feature of the KAFM model \cite{berthelemy2019local}. 
We argue that neither of these propositions is correct.
Firstly, this peak is found to be strongly sample dependent (Fig.\,\ref{fig5}a).  
It is the strongest for the initially reported translucent single crystals, where it roughly corresponds to entropy release of about $0.1R\,{\rm ln}(2)$ \cite{puphal2017strong}, i.e., 10\% of all spin-1/2 degrees of freedom.
Although this peak is still present in our transparent single crystals, its magnitude is reduced by an order of magnitude, so that the corresponding entropy release corresponds to only $\sim$1\% of all the spins. 
Secondly, we find that an applied magnetic field suppresses this peak and shifts it to slightly lower temperatures, independent of the field direction (Fig.\,\ref{fig5}b, c).
This isotropic behavior is substantially different from the highly anisotropic behavior of the broad maximum above $T_N$.
Therefore, we propose the 2-K peak is of extrinsic origin and most probably comes from CuO inclusions.
In fact, the N\'eel temperature of CuO is well known for its strong grain-size dependence when the dimension of the grains is reduced to a few nanometers -- it decreases from the bulk value $T_N = 229$\,K to $\sim$100\,K for 12-nm nanoparticles \cite{lepeshev2017particularities}, 30\,K for 5-nm nanoparticles and 13\,K when the particle size is further reduced to 2--3\,nm \cite{zheng2005finite}.
In our single crystals the peak appears at 2.0\,K, compared to 2.2\,K in Puphal's crystals \cite{puphal2017strong}, suggesting that the CuO inclusions are even smaller and much less abundant in transparent single crystals, making such impurities unobservable by ordinary optical methods and diffraction techniques.
Furthermore, as the magnetization in the magnetically ordered state of CuO nanoparticles, $M/H \simeq 0.02$\,cm$^3$/mol\,Cu  \cite{lepeshev2017particularities,zheng2005finite}, is weaker then the signal of \Cud~(Fig.\,\ref{fig4}b), CuO impurity fractions of the order of a few percents are also not observable in bulk magnetization measurements.      

\begin{figure*}[t]
\includegraphics[trim = 0mm 0mm 0mm 0mm, clip, width=1\linewidth]{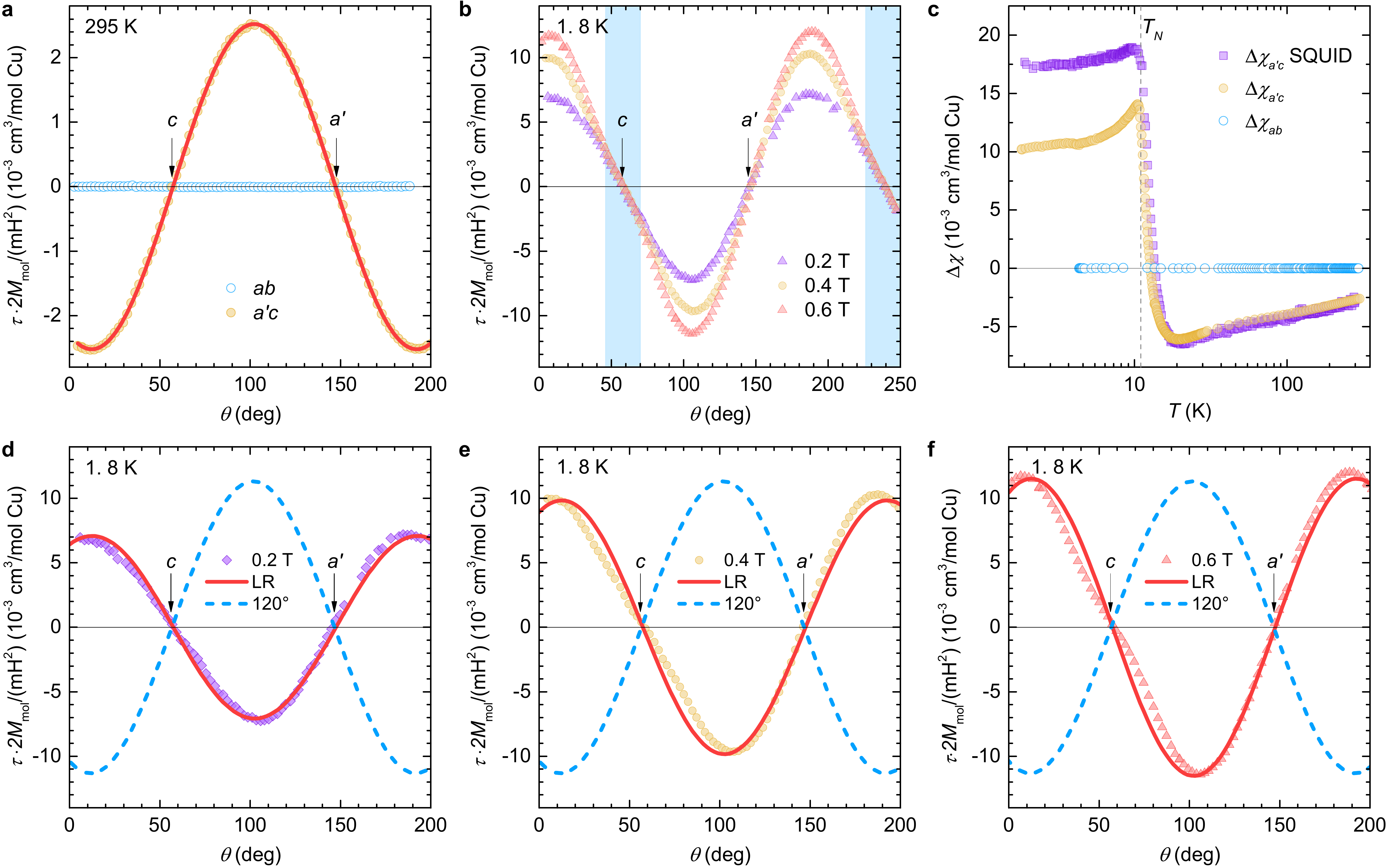}
\caption{The angular dependence of the magnetic torque $\tau$ in \Cud~normalize by squared applied magnetic field (a) in an applied field of 0.4\,T at 295\,K for two selected crystallographic planes, and (b) at 1.8\,K in several fields for the crystallographic plane containing the $c$ axis and an arbitrary $a'$ axis within the $ab$ kagome plane. 
The experimental data in (a) follow the linear-response (LR) model of Eq.\,(\ref{eq_torque}) shown by the thick solid line, while in (b) the 
measurements are consistent with the LR model, i.e., field independent, only in vicinity of the $c$ direction (blue shaded regions).
(c) The temperature dependence of the magnetic anisotropy $\Delta \chi$ in the $ab$ and $a'c$ planes from magnetic torque (at 0.4\,T) is compared to bulk SQUID measurements (at 1\,T).
The vertical line highlights the N\'eel temperature $T_N = 11$\,K.
(d--f) The 1.8-K torque measurements are compared to the LR model (thick solid lines) and a model of 120$^\circ$ coplanar order within the $ab$ plane, as found in \Cu~\cite{zorko2019negative} (dashed lines).
For the latter model the magnetic torque curves were calculated by assuming three spin sublattices using the approach of minimization of free energy \cite{herak2013magnetic}. 
The exchange interaction $J=82$\,K and out-of-plane Dzyaloshinskii-Moriya anisotropy $D/J=0.25$ were taken from the sibling compound \Cu~\cite{arh2020origin}.
} 
\label{fig6}
\end{figure*} 

\subsection{Magnetic Torque}
We have performed magnetic torque measurements in order to further characterize the magnetic order in \Cud.
These measurements were done on a custom-made apparatus using the torsion of a thin quartz fiber as a measure of the torque exerted on the sample by the applied magnetic field. 
The sample holder was made of ultra pure quartz and the data were collected in applied magnetic fields up to 0.8\,T at temperatures between 1.8 and 300\,K. 
Magnetic torque is a direct probe of magnetic anisotropy, which 
typically exhibits drastic changes due to symmetry reduction induced either by lattice distortions \cite{zorko2017symmetry} or long-range magnetic ordering \cite{herak2013magnetic}.
In the linear-response regime, i.e., for small applied magnetic fields $\mu_0H\ll k_B J/(g \mu_B)$ in the paramagnetic state, the magnetic torque is given by \cite{newnham2005properties}
\begin{equation}
\tau=\frac{m}{2M_{mol}} \mu_0 H^2\Delta\chi \sin (2\theta-2\theta_0),
\label{eq_torque}
\end{equation}
where $m$ is the sample mass, $M_{mol}$ its molar mass, and $\Delta\chi$ represents the susceptibility anisotropy in the plane within which the torque is measured by rotating the magnetic field through the angle $\theta$.  
Here, $\theta_0$ corresponds to the direction with maximal susceptibility within this plane.
It should be noted that Eq.\,(\ref{eq_torque}) also describes well  the magnetic torque in a single-domain collinear antiferromagnet, but only when the applied field is much smaller than the spin-flop field  \cite{herak2013magnetic}.  
In fields comparable to the spin-flop field strong deviations from Eq.\,(\ref{eq_torque}) are typically observed \cite{novosel2019strong}.

As expected, because of the trigonal crystal symmetry with the $c$ direction representing the high-symmetry three-fold rotational axis, we observe zero torque, i.e., zero magnetic anisotropy, within the crystallographic $ab$ plane at room temperature (Fig.\,\ref{fig6}a).
On the other hand, a pronounced anisotropy is found in the plane containing the $c$ axis and a direction within the $ab$ plane (denoted by $a'$ in Fig.\,\ref{fig6}).
At high temperatures, the linear-response model of Eq.\,(\ref{eq_torque}) fits the experiment very well (Fig.\,\ref{fig6}a) and yields $\chi_c > \chi_{ab}$, which agrees with bulk magnetic measurements at high temperatures (Fig.\,\ref{fig4}).
For $H||c$ the magnetic torque is zero, which is again in agreement with bulk symmetry requirements.
The temperature dependence of the susceptibility anisotropy $\Delta\chi_{a'c}=\chi_{ab} - \chi_c$ obtained from magnetic torque measurements in a field of 0.4\,T agrees well with bulk susceptibility measurements above $T_N=11$\,K (Fig.\,\ref{fig6}c).
It reaches a minimum around 20\,K, crosses zero and changes sign at 14\,K, and exhibits a maximum at $T_N=11$\,K.
The two curves, however, depart below $T_N$ because of the non-linear magnetic response, i.e., strong field-dependence of magnetic anisotropy below $T_N$, as explained below.   

The torque curves retain the $180^\circ$-periodicity in the $a'c$ plane  even below $T_N$ (Fig.\,\ref{fig6}b), however, the linear-response model of Eq.\,(\ref{eq_torque}) fits the experiment considerably worse than above $T_N$  (Fig.\,\ref{fig6}d-f).
The obvious deviation of experimental curves from a perfect sinusoidal dependence characteristic of a paramagnetic state is a clear signature of magnetic ordering.
Another experimental peculiarity that is a signature of magnetic ordering is that the susceptibility anisotropy -- the torque normalized by the applied field squared -- increases with the field (Fig.\,\ref{fig6}b).
This is in sharp contrast to the expected response of a paramagnet where the susceptibility anisotropy is constant in the low-field linear-response regime and starts decreasing at higher fields and low temperatures due to spin saturation effects.
The observed increase of the susceptibility anisotropy with an increasing applied field is thus rather unusual.
We note, though, that such behavior is consistent with bulk magnetic measurements, which show a clear maximum in susceptibility anisotropy around $\sim$1\,T (inset in Fig.\,\ref{fig4}c).
This is predominantly a consequence of the strong and non-monotonic field dependence of $\chi_{ab}$, while $\chi_{c}$ changes much more modestly with applied field (Fig.\,\ref{fig4}b).
Essentially, this is also the reason for the peculiar asymmetric shape of the torque curves at low temperatures  (Fig.\,\ref{fig6}b).
Namely, in a narrow angular region close to the $c$ direction the field-normalized torque curves measured at different fields actually overlap (see blue-shaded regions in Fig.\,\ref{fig6}b) while a strong field dependence is found for other directions.
As a result, the experimental curves become non-sinusoidal and non-symmetric at their maxima/minima, in sharp contrast to the linear-response model of Eq.\,(\ref{eq_torque}).
We note that the crystallographic $c$ axis remains a magnetic eigenaxis below $T_N$, as the torque curve always crosses zero at $H||c$ (Fig.\,\ref{fig6}b).
Moreover, within the experimental uncertainty the torque amplitude remains zero in the crystallographic $ab$ plane down to the lowest temperatures.
Therefore, bulk three-fold symmetry remains preserved even in the magnetically ordered state of \Cud.

\section{Discussion}

The optimization of the hydrothermal synthesis route has led to the growth of transparent single crystals of \Cud~with much better purity compared to  previously available crystals and powders \cite{puphal2017strong, berthelemy2019local}.
Contrary to previous samples, our crystals lack any detectable amount of clinoatacamite impurities, while CuO impurities are also reduced below the XRD detection threshold but are nevertheless likely responsible for the sample-dependent small peak observed in specific heat at 2\,K.   
The level of quality of these crystals might be related to the Y-site disorder, which seems to be present in our crystals but was not reported in previous studies \cite{puphal2017strong, berthelemy2019local}. 
In this respect we note that atomic scale imperfections, such as point defects, dislocations and disorder, have been reported to promote the growth of large single crystals with good optical qualities such as transparency and clarity \cite{sun2012investigation} while larger nano- and micro-sized inclusions usually degrade the optical quality of crystals.

Contrary to previous claims of a spin-liquid ground state in \Cud~\cite{berthelemy2019local}, we find that high-purity single crystals actually undergo a bulk magnetic-ordering transition at $T_N=11$\,K.
Therefore, we propose that the lack of detectable magnetic ordering in less pure \Cud~samples is due to structural disorder.
In fact, a similar situation was previously encountered in \Cu, where a heterogeneous partially frozen magnetic ground state was detected in samples containing impurity phases \cite{berthelemy2019local}, while in pure samples a homogeneous long-range ordered magnetic state was discovered \cite{zorko2019YCu3muon}.
Moreover, in another KAFM material vesignieite, Cu$_3$Ba(VO$_5$H)$_2$, coexisting frozen magnetic moments and fluctuating moments were also initially reported for powder samples \cite{colman2011spin}, but a fully ordered magnetic state was later found in purer single crystals \cite{yoshida2012magnetic}.
The suppression of magnetic order by non-optimal crystallization in these systems can be understood within the scenario of disorder-induced bond randomness leading to spin-liquid-like random singlet states \cite{kimchi2018valence} on the kagome lattice \cite{kawamura2014quantum,shimokawa2015static,kawamura2019nature}.
Contrary to conventional magnetically ordered states, where bond randomness is irrelevant, such randomness should be much more relevant on frustrated lattices with many competing phases.  
The magnetic order found in our pure \Cud~crystals is likely unusual and coexists with persistent spin dynamics, as it was found also in \Cu~\cite{zorko2019YCu3muon}.

Our experiments have revealed that the applied magnetic field influences the magnetically ordered state of pure \Cud~single crystals in a non-linear and strongly anisotropic manner (Figs.\,\ref{fig4}b and \ref{fig6}b).
It leads to a pronounced non-monotonic effect when it is applied within the kagome planes and a much weaker effect when it is applied perpendicular to the kagome planes, which can only be attributed to non-trivial magnetic order.     
Before trying to determine the specifics of this order,
let us note that a rather small ratio of transparent single crystals  compared to all crystals resulting from our synthesis unfortunately does not allow us to produce large enough quantities of pure single crystals  for conventional magnetic neutron diffraction experiments.
Nevertheless, the magnetic torque measurements below $T_N$ provide an informative alternative insight into the magnetic ground state.

These measurement first reveal that there is no ferromagnetic component to the long-range order, as this would yield a torque contribution with $360^\circ$ periodicity, $\tau_0^{FM} \sin (\theta-\theta_0^{FM})$, which is not observed (Fig.\,\ref{fig6}).
Simple models of antiferromagnetic collinear spin structures also don't agree with the experiment.
The first such model, where the spins are confined to the $ab$ plane, could provide net zero magnetic torque within this plane assuming three equally populated magnetic domains.
However, this model requires $c$ to be the magnetic hard axis and therefore leads to $\chi_c > \chi_{ab}$, which is not consistent with experiment (Fig.\,\ref{fig4}b).
On the other hand, the experimentally determined $\chi_c < \chi_{ab}$ below $T_N$ is reproduced by collinear long-range order along the $c$ axis. 
However, in sharp contrast to experiment, the field dependence of susceptibility should be in this case much more pronounced along the easy axis ($c$ direction) than perpendicular to it when the applied magnetic field starts inducing non-linear effects such as spin flop  \cite{novosel2019strong}.
Turning next to non-collinear orders, which are actually preferred by frustrated spin lattices, we examine the 120$^\circ$ ($q=0$) coplanar spin structure that was observed in the sibling compound \Cu~by neutron diffraction \cite{zorko2019negative}.
We find that this order also does not comply with the experiment. 
It yields a field-independent susceptibility anisotropy (dashed lines in Figs.\,\ref{fig6}d-f) and, moreover, misses the proper phase of the torque curves by 180$^\circ$, as it predicts $\chi_c > \chi_{ab}$, just like the model of in-plane collinear order.  

The magnetic order realized in pure crystals of \Cud~is thus apparently more complicated than the chiral $q=0$ order reported for \Cu~\cite{zorko2019negative}. 
This is not surprising, because even though the two crystal structures are similar, the unit cell of the distorted kagome lattice in \Cud~comprises nine spins (Fig.\,\ref{fig3}c), while only three spins exist in the unit cell of the more symmetric \Cu.
Due to the distortion and consequently two inequivalent magnetic Cu$^{2+}$ sites, the single nearest-neighbor distance of 3.372\,\AA~found in \Cu~is replaced by three different distances: $d_1 = 3.256$\,\AA, $d_2=3.371$\,\AA, and $d_3=3.374$\,\AA~(Fig.\,\ref{fig3}c).
This likely affects the nearest-neighbor exchange profoundly and leads to a spatially anisotropic lattice with three different nearest-neighbor exchange interactions and, consequently, a different ordered state than found in \Cu.
According to spin-wave calculations, the $q=0$ three-spin 120$^\circ$ order of the symmetric kagome lattice is destabilized by spatially anisotropic exchange interactions on the basic kagome spin triangle and only large magnetic anisotropy can render this structure stable \cite{coker2021effects}.
Although nothing is known in literature about the state that supersedes the 120$^\circ$ state on a spatially anisotropic kagome lattice like the one found in \Cud, we would expect a magnetically ordered state with a larger unit cell to be realized rather than a quantum spin liquid, as lowering the symmetry of the KAFM Hamiltonian in general relieves frustration and reduces the classical degeneracy of the ground state \cite{yavors2007heisenberg, wang2007quantum}, i.e., reduces the system's tendency towards quantum disorder.

\section{Conclusions}
Optimization of the hydrothermal synthesis has allowed us to grow high-purity single crystals of the distorted KAFM material \Cud.
Hand-picked transparent single crystals lack dark inclusions, in sharp contrast to translucent crystals and aggregates that are also produced as well as to all previously reported single-crystal and powder samples, all suffering from parasitic clinoatacamite and CuO impurities \cite{puphal2017strong, berthelemy2019local}.
Our experimental investigation combining bulk magnetic susceptibility, specific heat and magnetic torque measurements has revealed, unexpectedly, that these pure single crystals undergo magnetic ordering below $T_N=11$~K, while the lack of magnetic ordering in less pure \Cud~single crystals and powders is attributed to impurities and not to the intrinsic distortion of the crystal lattice.
The magnetic ordering is thus much alike the ordering of the sibling compound \Cu~with a perfectly symmetric kagome lattice.
However, the simple three-spin 120$^\circ$ magnetic order observed in the latter compound is not consistent with the non-monotonic and highly asymmetric experimental magnetic-torque curves.
Therefore, the long-range ordered spin pattern in \Cud~is likely more complicated.
It would be highly relevant for future studies to determine to what extend the isotropic KAFM model is perturbed in this compound and to exactly determine its magnetically ordered ground state.
Acquiring a better understanding of the effects of lattice distortions in KAFM materials is of highest importance \cite{norman2019valence}, because these may even appear spontaneously to relieve frustration in the ground state, as recently suggested for the archetypal KAFM compound herbertsmithite \cite{zorko2017symmetry}.   

\acknowledgments{We acknowledge the financial support of the Slovenian Research Agency through the Program No.~P1-0125 and Projects No.~N1-0148, J1-2461, and Z1-1852.}

%

\end{document}